\documentclass[%
 reprint,
showpacs,
showkeys,
 amsmath,amssymb,
 aps,
]{revtex4-1}
\usepackage[normalem]{ulem}
\usepackage{bbold}
\usepackage{float}
\usepackage{graphicx}
\usepackage{dcolumn}
\usepackage{bm}

\usepackage{relsize}
\usepackage{color}

\begin{document}

\preprint{APS/123-QED}

\title{Non-locality in high spin systems with tensor correlations}

\author{V. Ravishankar}
 \altaffiliation{Department of Physics, Indian Institute of Technology, Delhi}
\author{R. P. Sandhir}%
 \altaffiliation{Department of Physics and Computer Science, Dayalbagh Educational Institute, Agra}

\begin{abstract}
We address the problem of detecting non-locality in coupled $N$ level systems in the language of spin.  Through a number of examples, we show that non-locality can be detected via a violation of the standard Bell inequality, irrespective of what $N$ is, with correlations from observables of order $k \sim 2s$ in the fundamental spin operators. We further show that, contrarily, if the order $k$ is frozen, then non-locality eludes a detection when  $s \gg k$, leading to a weak classical limit. Armed with these results, we  proceed to characterize observables that `genuinely' reflect non-locality in higher dimensions, and demonstrate that one needs to go beyond the standard set-ups such as Stern-Gerlach and those with correlations involving measurement of spin projections along quantization axes. 
\end{abstract}

\pacs{03.65.Ud, 03.65.Aa, 03.67.Mn}
\keywords{Bell Inequality, Classical Limit, N-level Systems, Non-locality, Spin-Spin Correlations}
\maketitle


\section{Introduction}\label{sec:Introduction}
There exists a fairly large  body of work on non-locality in  $N$ level systems\cite{Bellinequalities,Merm80,Kasz00,Garg84,Peres92,Brau92,Gis92,Coll02,Wodk95,Wu01} which explore non-locality mainly from the view point of identifying appropriate correlations, and to explore possible attainment of classical limit in the large $N$ limit. Generalizations to multipartite systems have been accomplished by construction of correlation functions for higher dimensions \cite{FuLi04} and by establishing multi-party Bell inequalities. Important that these questions are, we argue that they need refinement before we look for answers. Indeed, if no restrictions are imposed on $N$ level systems, discovering non-locality is trivial in the following sense \cite{Bellinequalities}: One merely needs to identify a four dimensional subspace spanned by four separable states, and the rays in the subspace mimic a two-qubit state. This  leads to a maximal violation of the Bell inequality, which cannot be improved upon, thanks to the Cirel'son bound \cite{Cirelson80}. Similar arguments can be advanced for other measures of inequalities which involve more complicated inequalities with larger number of correlations. The refinement that is required is the stipulation that the correlations genuinely probe the Hilbert space fully and not be trapped in a subspace. This can considerably complicate the analysis, as reflected in the divergent conclusions drawn in literature.

The purpose of this paper is to revisit non-locality in $N$ level systems keeping the stipulation mentioned above in mind, with exclusive emphasis on spin systems.  High spin systems are ubiquitous, and spin dependent interactions make the preparation and manipulation of states relatively easier. 
In contrast, manipulation of  multiqubit states to prepare $N$ level systems, and then couple them to form a bipartite state is much more daunting though generalised Svetlichny's inequalities have been proposed for the same \cite{Coll02_2,Mitch04}.  In a similar manner, measurements of correlations also gets complicated. High spin systems do not suffer from these drawbacks.
Further, correlations in high spin systems  can be studied through observables which are of different orders in the fundamental spin operators. One knows from atomic and nuclear physics that measurement of observables which are of higher order in spin is more difficult than those of lower orders. With these in mind, we investigate (i) the `simplest' observables that lead to the largest possible non-locality, (ii) their behaviour as $N \rightarrow \infty$, and (iii) the precise import of the so called quantum-classical transition in that limit. Rather than trying to prove general theorems, we consider representative examples to illustrate the general trend. For our purposes, it suffices to consider observables which are upto quartic degree in spin. We make  a rather detailed comparison with existing results, wherever possible.

\section{Preliminaries and the Experimental Setting}
\begin{enumerate}
\item Throughout the paper, we   consider a bipartite system of two equal spins, $s$. We invariably study non-locality in  one representative fully entangled (Bell) state,  {\it viz}, the singlet state given by
\begin{equation}
 |0_s> =\frac{(-1)^{s}}{\sqrt{2s+1}}\sum_{m=-s}^{s}(-1)^{m}|m,-m>.
 \end{equation}  
The notation in the ket  in the LHS emphasizes that the state is isotropic, with total spin zero. The isotropy makes the choice of correlations simpler.
\item  We employ the Bell inequality in its standard form, formulated for the Bell function:
\begin{equation}
\mathcal{B} \equiv |\mathcal{C}(a, b) - \mathcal{C}(a, b')|+|\mathcal{C}(a',b)+\mathcal{C}(a',b')|,
\label{eq: bellinequality}
\end{equation}
where the spin-spin correlation \newline
$C(a,b) = \langle 0_s\vert O_A(a)O_B(b) \vert 0_s\rangle$ is defined in terms of the expectation values of products of two observables, each belonging to a subsystem, as indicated. It should be emphasized that the arguments $\{a,~b,~a',~b'\}$ refer to collective variables that constitute the parameter space. Indeed, the state exhibits non-locality if the Bell function violates the inequality ${\mathcal B} \le 2$, i.e., there are some regions in the parameter space where ${\cal B} > 2$. 
\item We shall denote by $O^s_k(a)$,  observables of order $k$ in the spin operators, for a spin $s$ particle. Similarly, we  denote by $C^s_k(a,b) $, a correlation constructed from two observables of the same order $k$ in spin operators. The corresponding Bell function will be denoted by ${\cal B}^s_k$ \footnote{More generally, one could consider correlations of degree $k~(k')$ in subsystem $A(B)$, but they are of no consequence to us here.}.
\item Customarily, and especially in experiments, the parameters are taken to be quantization axes with respect to  the spin observables measured. We adhere to the same tradition here. 
Furthermore, the experimental configuration considered throughout the paper is the standard planar configuration depicted in Fig.~\ref{fig:planar_geometry}, with the configuration:
\begin{equation}
\theta_{ab}=\theta_{a'b}=\theta_{a'b'}=\frac{\theta_{ab'}}{3},
\label{eq:angles}
\end{equation}
in which the violation of the Bell Inequality, if any, is always maximal.  We may note parenthetically that this choice of the parameters is not general enough for spins $\ge \frac{1}{2}$, the importance of which we briefly discuss in the concluding section.

\begin{figure}[H]
\centering
\includegraphics[width=0.3\textwidth]{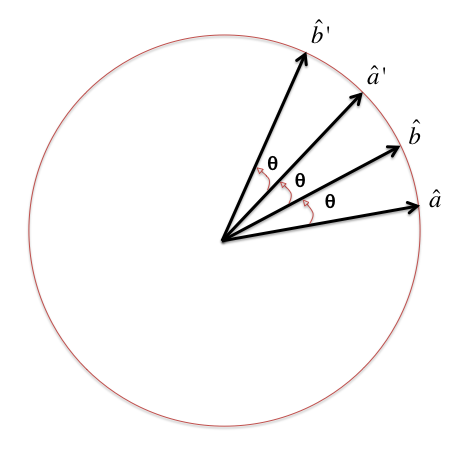}
\caption{The standard experimental configuration considered for Bell-type experiments involving quantisation axes $\hat{a},\hat{a'},\hat{b}$ and $\hat{b'}$, with angles defined by Eq.~\ref{eq:angles}.}
\label{fig:planar_geometry}
\end{figure}

\end{enumerate}
 In the simplest and the most studied two qubit case, the choice of the correlation  is essentially unique, and is given by $C = (\vec{\sigma}_A\cdot \hat{a})(\vec{\sigma}_B\cdot \hat{b})$. The parameter spaces are the respective Bloch spheres, and  the inequality is violated by all entangled pure states, and maximally by a Bell state. The violation is also the maximum allowed for a quantum system with this particular Bell formulation, with the Bell function taking the value $2\sqrt{2}$.

\section{Bell inequality in higher spin states: Linear correlations}\label{sect:vector}
The number of linearly independent correlations for a spin $s$ state is $O(N^4)$, but  the Bell function does not follow linearity in correlations, and that is the crux of the problem. It is, therefore, advisable to consider specific examples to start with. Consider the linear correlation 
\begin{equation}
\mathcal{C}_{1}^s(\hat{a}, \hat{b}) =  \langle 0_s\vert (\vec{\Sigma}^s_A\cdot \hat{a}) (\vec{\Sigma}^s_B\cdot \hat{b}) \vert 0_s \rangle; ~~
\vec{\Sigma}^s_{A,B}  =  \frac{\vec{S}_{A,B}}{s}
\label{eq:dipole}
\end{equation}
in terms of the normalized spin operators $\vec{\Sigma}^s$, which  respect the unit norm bound: they satisfy $\vert Tr\{\rho^s \Sigma^s_i \}\vert \le 1$ for all states.
Let $\cos \theta_{ab} \equiv \hat{a}\cdot \hat{b}$. Exploiting bilinearity in  the spin operators and the isotropy of the state,  we find
\begin{equation}
\mathcal{C}^s_1(\hat{a}, \hat{b}) = -\frac{s+1}{3s} \cos \theta_{ab} \equiv \frac{s+1}{3s}\mathcal{C}_1^{\frac{1}{2}}(\hat{a}, \hat{b}).
\label{eq:linear_corr}
\end{equation}
Note that the spin dependent scaling factor, which is also inherited by the corresponding Bell function ${\cal B}^s_1$,  decreases rapidly with $s$, achieving an asymptotic value  of $\frac{1}{3}$. It falls by a factor 2/3 for $s=1$. Since the maximum violation for two qubit case is $2\sqrt{2}$, it follows that Bell inequality is respected by the correlation for spins $s \ge 1$. So one is obliged to look at  correlations which involve higher orders in spin.
\section{Biquadratic Correlations}
We consider the simplest case, $s=1$, first.
\subsection{s=1}\label{sect:tensor}
Since the linear correlations fail to reveal non-locality,  we  move on to  biquadratic correlations.  
It would appear that the choice $O= (\vec{\Sigma}^s\cdot \hat{a})^2$ is natural.  But that is not to be, since its expectation value is confined to the subinterval $[0,1]$, with the corresponding correlation  having the form $\frac{1}{3}(1 + \cos^2\theta)$, which peaks at $\frac{2}{3}$, which immediately negates the possibility of a violation.
In contrast,  consider the trial observable
\begin{equation}
O^{(2)}= 2(\vec{\Sigma}^1\cdot \hat{a})^2 -1
\label{eq:quad_s1}
\end{equation}
 whose  expectation values cover the full range $[-1,+1]$.
It is easy to verify that
\begin{eqnarray}
\mathcal{C}^{1}_{2} & =& \langle  0_1\vert \{2(\vec{\Sigma}^1_A\cdot \hat{a})^2 -1\}\{2(\vec{\Sigma}^1_B\cdot \hat{b})^2 -1\} \vert 0_1\rangle \nonumber \\
&=& \frac{1}{3}(4\cos^2 \theta-1)
\label{eq:quadcorr_s1}
\end{eqnarray}
We note that this correlation has a support in the much larger interval $[-\frac{1}{3}, +1]$, though it falls short of the support $[-1, +1]$ for the correlation in the two qubit case.   For this reason, one may anticipate that a violation of Bell inequality, if any, would be smaller than in the two qubit case. 

\begin{figure}[H]
\includegraphics[width=0.5\textwidth]{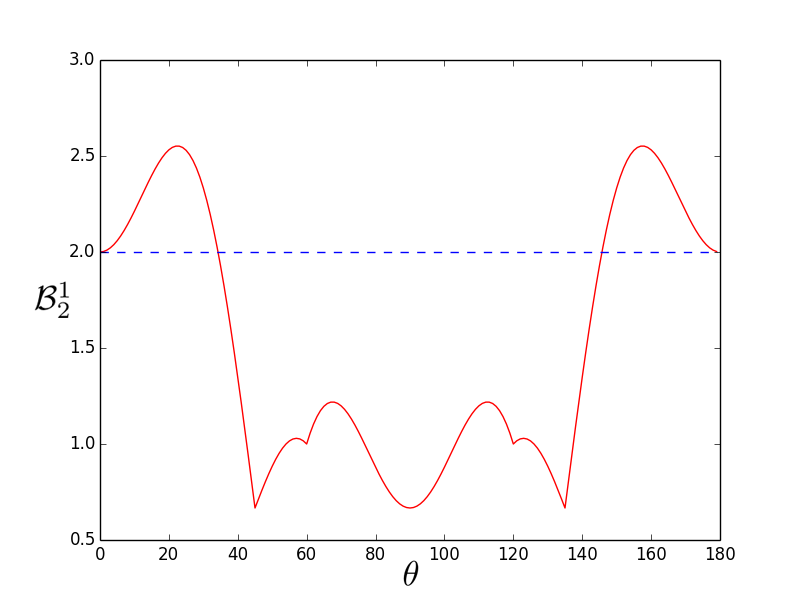}
\caption{Bell function $\mathcal{B}^1_2$ for the $spin-1$ singlet state using quadratic operators given by Eq.~\ref{eq:quad_s1}.}
\label{fig:bell_inequality_quadratic_corr}
\end{figure}

The corresponding Bell function is plotted in red in the standard planar geometry  (Fig.\ref{fig:bell_inequality_quadratic_corr}). There is a clear violation of the inequality in the  sector $[0.97, 2.17]$ {\it rad}. The maximum value attained is 2.55, and the percentage area of violation is found to be about 14.7\%. These numbers may be contrasted with the two qubit case which shows violation in the intervals $[0.1,0.19]$ and  $[1.95,\pi]~ rad$ with the maximum value $2\sqrt{2} \approx 2.82$, and a  total percentage area of  16.32\%. We return to a critique of these results later.

\subsubsection{Experimental determination of $C_{2}^1$}
We briefly digress  to discuss how $C^1_2(\hat{a},\hat{b})$ may be determined  experimentally through appropriate count rate measurements.  Observe that $\Pi_0(\hat{a} )\equiv (1- (\vec{S}\cdot \hat{a})^2)$,  is the projection operator for the state $m=0$ along the quantization axis $\hat{a}$.  The correlation can be written as:
\begin{eqnarray}
\mathcal{C}^1_{2} & = & \langle 0_1 \vert \{1-2\Pi^A_0(\hat{a})\}\{1-2\Pi^B_0(\hat{b})\} \vert 0_1 \rangle \nonumber \\
& = & -\frac{1}{3} + 4\langle 0_1 \vert \Pi^{AB}_0 (\hat{a}, \hat{b}) \vert 0_1\rangle \nonumber \\
& = & -\frac{1}{3} + \frac{4N^{AB}_0(\hat{a},\hat{b})}{N}
\end{eqnarray}
where the last line gives the measurement prescription explicitly, in terms of $N^{AB}_0(\hat{a},\hat{b})$ which  is the joint count rate for the two spins to be in the state $m=0$ along their respective quantization axes, and $N$ is the total count rate.
\subsection{Quadratic correlations for $s \ge 3/2$}
We first note that 
\begin{equation}
\langle 0_s\vert (\vec{\Sigma}^s_A\cdot\hat{a})^2 (\vec{\Sigma}^s_B \cdot \hat{b})^2\vert 0_s\rangle  = 
F(s)  + G(s)\cos^2\theta 
\end{equation}
where the scaling factors,
\begin{eqnarray}
F(s) &  = & \frac{2s^3 + 4s^2 +3s +1}{30s^3} \nonumber \\
G(s) & =  &  \frac{4s^3 +8s^2 +s -3}{30s^3},
\label{eq:FsGs}
\end{eqnarray}
\begin{figure}[H]
\includegraphics[width=0.5\textwidth]{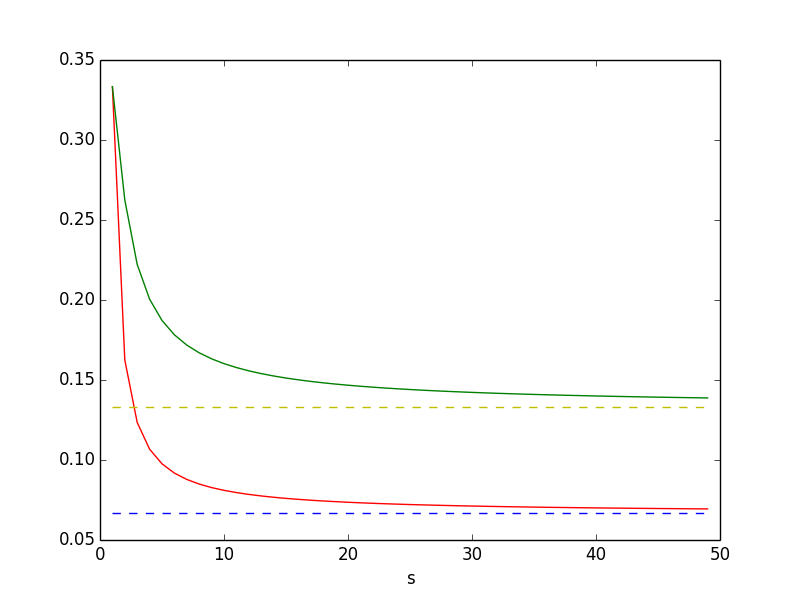}
\caption{Green: scaling factor $G(s)$, Red: scaling factor $F(s)$, Yellow: bound for scaling factor G(s) - $\frac
{2}{15}$, Blue: bound for scaling factor F(s) - $\frac{1}{15}$.}
\label{fig:FsGs}
\end{figure}

\noindent are monotonically decreasing functions of $s$, as may be seen in Fig. ~\ref{fig:FsGs},  and attain their asymptotic values $ F= 1/15,~ G =2/15$ in the limit $s \rightarrow \infty$.  

The  correlation $\mathcal{C}_{2}^s$, corresponding to $O_A=2(\vec{\Sigma}^s_A\cdot \hat{a})^2-1$ (and similarly for $O_B$), is evaluated to be 
\begin{equation}
\mathcal{C}_{2}^s = 4\{F(s) + G(s)\cos^2\theta\} - \frac{4}{3}\frac{s+1}{s} +1.
\label{eq:correlations2}
\end{equation}

\begin{figure}[H]
\includegraphics[width=0.5\textwidth]{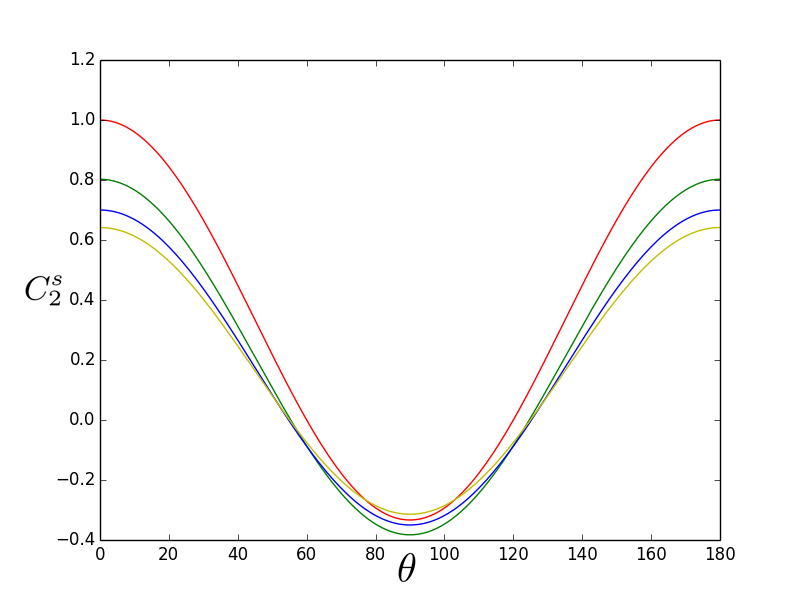}
\caption{Correlation $C_{2}^s$ given by Eq.~\ref{eq:correlations2} for Red: $s=1$, Green: $s=\frac{3}{2}$, Blue: $s=2$, Yellow: $s=\frac{5}{2}$.}
\label{fig:correlations2}
\end{figure}

\begin{figure}[H]
\includegraphics[width=0.5\textwidth]{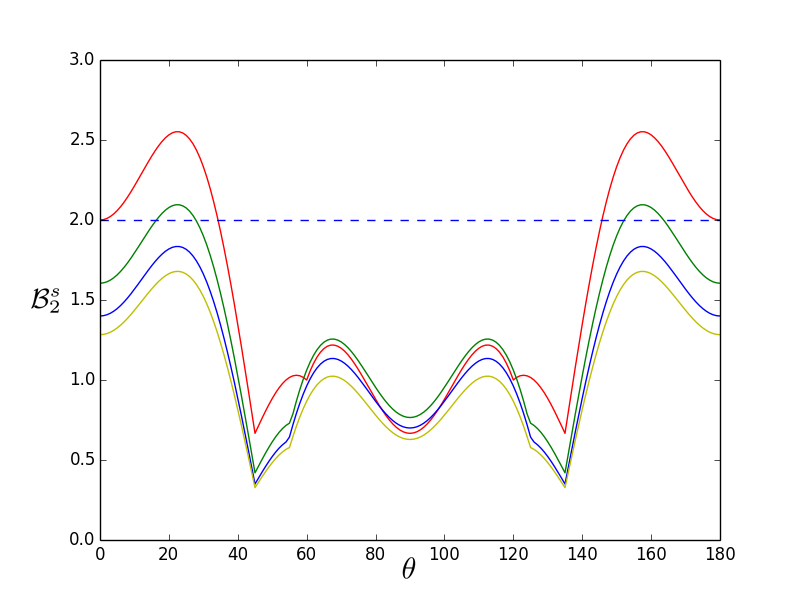}
\caption{Bell function $\mathcal{B}_{2}^s$ for the $spin-s$ singlet state using quadratic operators given by Eq.~\ref{eq:quad_s1}. Red: $s=1$, Green: $s=\frac{3}{2}$, Blue: $s=2$, Yellow: $s=\frac{5}{2}$, Blue dashed: bound of the inequality.}
\label{fig:bell_correlations2}
\end{figure}

Fig. \ref{fig:correlations2} depicts the behaviour of the correlation as a function of $\theta$ for various values of $s$. The range of the correlation is seen to diminish rapidly as $s$ is increased. The corresponding Bell functions ${\cal B}_{2}^s$ are plotted in Fig.~\ref{fig:bell_correlations2} in the standard planar configuration in the parameter space. It is clear that there is a very mild violation for $s=\frac{3}{2}$, with a maximum value of 2.09, and none whatsoever for $s \ge 2$. This raises the question whether the correlation chosen is optimal for all spins, and if 
higher order observables  are required when $s \ge 2$.  We seek to settle this issue through a global (numerical) search in the next section.

\section{Global study of biquadratic correlations}\label{sect:generic}
Our studies have so far been restricted to a few special examples which do not, however, shed complete light on the extent to which non-locality may be detected in higher spin systems. To rectify this drawback,  We  consider a general quadratic observable:
\begin{equation}
O  = C_2 (\vec{\Sigma}^s\cdot \hat{a})^2 + C_1 (\vec{\Sigma}^s\cdot \hat{a}) + C_0
\label{eq:quadg}
\end{equation}
with the proviso that its expectation values be bounded by unit norm. We fix  the optimal values of the constants $C_i$  through a numerical search\footnote{The numerical results agree with the analytic results which are easily available for $s=\frac{1}{2},~\frac{3}{2}$.}. The global search also determines whether several of the prescriptions given by the earlier works do really yield the maximum possible violations or not.
\begin{figure}[H]
\includegraphics[width=0.4\textwidth]{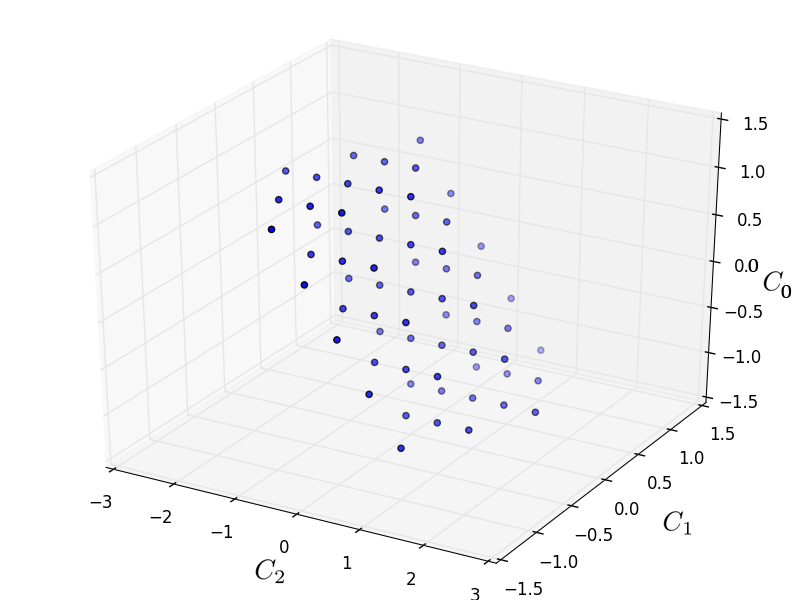}
\caption{Scatter plot of coefficients $C_0,C_1,C_2$ for $spin-1$ constraints given by Eq.~\ref{eq:constraints_quad} shown in steps of 0.5 for convenience.}
\label{fig:scatter_abc_s1}
\end{figure}
The unit norm bound on the coefficients $C_i$ in Eq. \ref{eq:quadg} simply translates to the the set of $2s+1$ constraints (in each $m$ value)
\begin{equation}
-1\leq \frac{C_2 m^2}{s^2}+\frac{C_1 m}{s}+C_0\leq 1 ~~\forall m \in [-s,s]
\label{eq:constraints_quad}
\end{equation}
 in the three parameters. The  data points for the coefficients are generated by selecting the bounds [-5,5] for each coefficient here and everywhere. Each of the equalities  yields a two dimensional plane in the $(2s+1)$ dimensional space, and the intersections of the planes gives us the vertices of a polyhedron within and on which the coefficients $C_i$ are constrained to lie. The additional requirement that the observables admit their maximum value restricts our search to identifying one of the vertices. We present the results for various spins in the following subsections.
\subsection{ $s=1$}
\begin{figure}[H]
\includegraphics[width=0.5\textwidth]{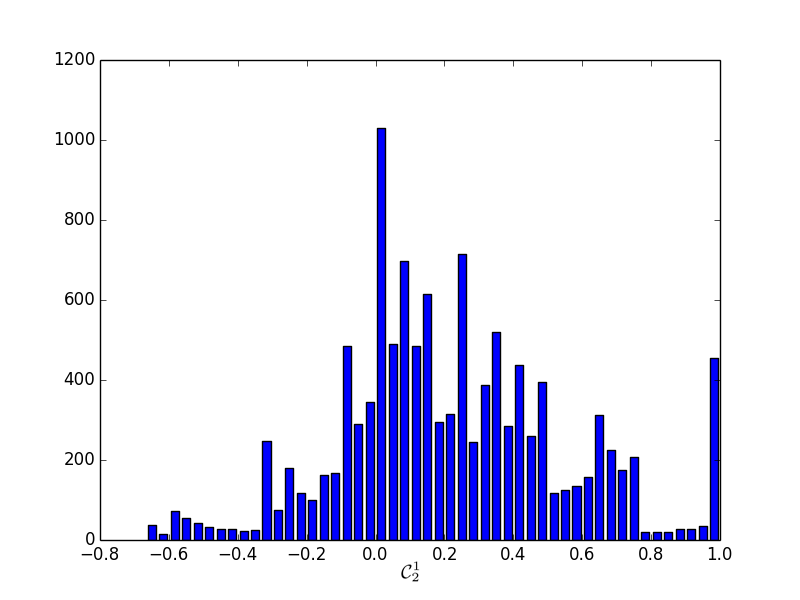}
\caption{Correlator $\mathcal{C}_2^1$ for all possible valid coefficient and configuration combinations.}
\label{fig:hist_corr_quadratic_s1}
\end{figure}

\begin{figure}[H]
\includegraphics[width=0.5\textwidth]{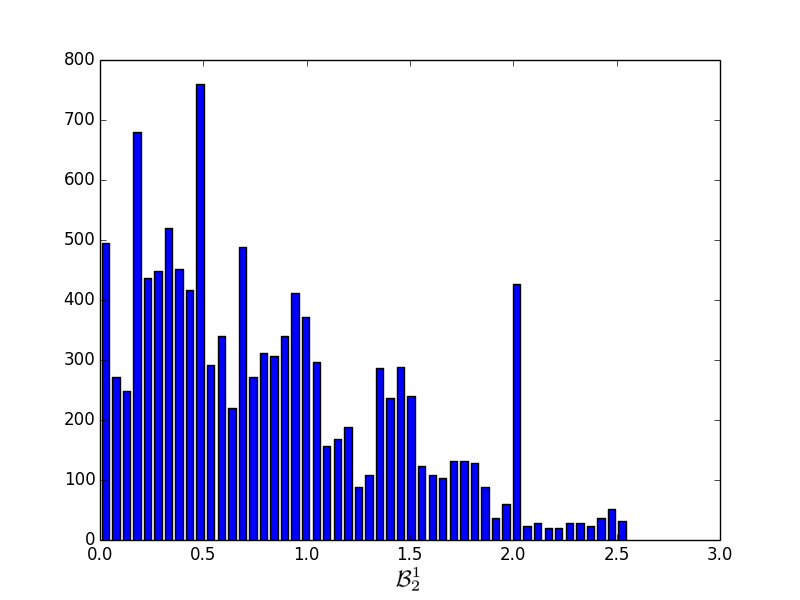}
\caption{Bell function $\mathcal{B}^1_2$ for all possible valid coefficient and configuration combinations.}
\label{fig:hist_bell_quadratic_s1}
\end{figure}

  When $s=1$, we find that the maximum violation occurs when $C_2=2;~C_1=0;~C_0=-1$ which is exactly the trial observable that we constructed in the previous section.  The numerical search confirms that the maximum violation occurs in the planar geometry, with ${\cal B}_{max}=2.55$ as reported. More detailed results are presented in 
  the scatter plot in Fig. \ref{fig:scatter_abc_s1} which displays  the planar segment  in which the coefficients $C_i$ are constrained, and  
Fig.~\ref{fig:hist_corr_quadratic_s1} which  shows a histogram plot of the values of the correlation for all coefficients $C_i$ that are consistent with the constraints. The corresponding histogram for the Bell function, which is of direct interest, is shown in Fig.~\ref{fig:hist_bell_quadratic_s1}.  From this we estimate that the relative  region over  which the violation takes place is  $\sim 15\%$.  

In short, we may conclude that it is not possible to improve upon the violation seen for the correlation chosen in Eq.~\ref{eq:quad_s1}. The maximum violation, pegged at the value 2.55 falls short of the maximum allowed value, 2.88, which is realized for a two qubit system -- notwithstanding the fact that we are dealing with a completely entangled pure state. This conundrum will be addressed in the concluding section.

\subsection{  $s \ge \frac{3}{2}$}
\begin{figure}[H]
\includegraphics[width=0.5\textwidth]{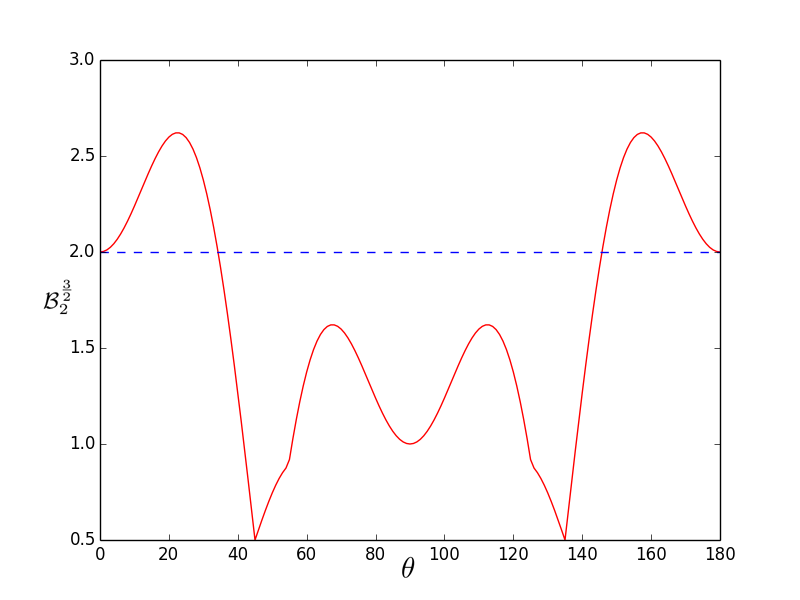}
\caption{Bell function $\mathcal{B}_2^{\frac{3}{2}}$ for the $spin-\frac{3}{2}$ singlet state using quadratic operators given by Eq.~\ref{eq:gen_quad_op_s1p5}. Blue dashed: bound of the inequality.}
\label{fig:bell_gen_1p5}
\end{figure}

 Consider $s=\frac{3}{2}$ first.  The optimal  values of the coefficients are found to be   $C_2=2.25, C_1=0,C_0=-1.25$, yielding the observable
\begin{equation}
O_A=\frac{9}{4}(\vec{\Sigma}_A\cdot \hat{a})^2 -\frac{5}{4}
\label{eq:gen_quad_op_s1p5}
\end{equation}
which can be recast into the elegant form    
\begin{equation}
O_A (\hat{a}) = \Pi_{3/2}(\hat{a})  + \Pi_{-3/2}(\hat{a}) - \Pi_{1/2}(\hat{a})- \Pi_{1/2}(\hat{a}) 
\label{eq:gen_quad_s1p5_op}
\end{equation}
in terms of the projection operators $\Pi_m \equiv \vert m \rangle \langle m \vert$ along the quantization axis $\hat{a}$,
with a similar expression for $O_B$. Note that expectation values of $O_{A,B}$ span the full range 
$[-1,+1]$. The correlation also assumes the elegant form
\begin{equation}
\mathcal{C}_2^{3/2} = P_2(\cos\theta)
\end{equation}
which has its support in $[-\frac{1}{2}, 1]$, larger than the one obtained for $s=1$. The Bell function plotted in Fig.~\ref{fig:bell_gen_1p5} clearly shows that non-locality is exhibited by these correlations. The maximum value is 2.62, which is larger than the violation for $s=1$. The  percentage area of violation is given by $\approx 7.85\%$ which, however, is smaller than the case for $s=1$. Incidentally, 
note that the coincidence count rates to be measured are clear from the very expression for  the observables given in Eqn. \ref{eq:gen_quad_s1p5_op}.

The increase in the violation from $s=1$ to $s=\frac{3}{2}$ is anomalous. In fact, we find that the biquadratic correlation fails to violate Bell inequality everywhere in the parameter space if $s \ge 2$.

It is getting clear from the studies so far that violation of the Bell inequality requires correlations of observables which are of maximal order in the spin variables. To strengthen this qualitative conclusion, and also to draw more quantitative conclusions for comparison with similar findings, and claims made so far \cite{Merm80, Gis92,Wodk95,Bana98,Peres92}, we consider two more cases -- generic  observables of degree three, and a specific observable of degree four.  This necessitates a discussion of $s=2$ states as well. We address the cubic case first, and conduct a global search in the parameter space.

\section{ Global study of Cubic Correlations }
The generic form of the observable,  say for the subsystem $A$, is given by
\begin{align}
O_A=C_3(\vec{\Sigma}^s_A\cdot a)^3+C_2(\vec{\Sigma}^s_A\cdot a)^2+C_1(\vec{\Sigma}^s_A\cdot a)+C_0.
\end{align}
The usual requirement that $O_A$ be bounded by unit norm leads to the set of  $2s+1$ constraints
\begin{equation}
 -1 \le \frac{C_3 m^3}{s^3}+\frac{C_2 m^2}{s^2}+\frac{C_1 m}{s}+C_0 \le1~ \forall m \in [-s,s].
\label{eq:constraints_cubic}
 \end{equation}
As before, the coefficient data were generated by selecting the bounds [-5,5]. The search in the parameter space yields a maxima in violation of Bell inequality in two orthogonal subspaces,  of even and odd parity in spin observables. The even parity case has already been disposed off in the previous section.  We address the complementary space. 
\subsubsection{$s=\frac{3}{2}$}
\begin{figure}[htp]
\includegraphics[width=0.5\textwidth]{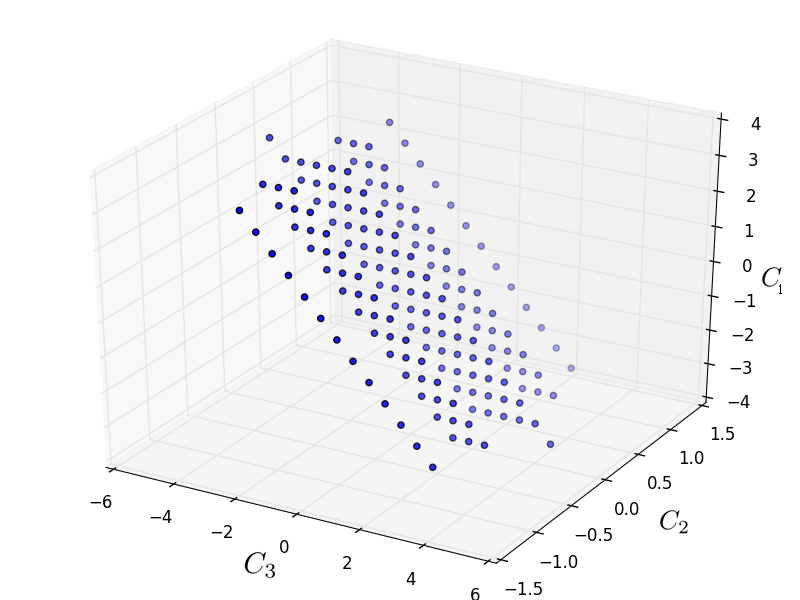}
\caption{Scatter plot of coefficients $C_1,C_2,C_3$ for $C_0=0$ for $spin-\frac{3}{2}$ constraints given by Eq.~\ref{eq:constraints_cubic}
shown in steps of 0.5 for convenience.} 
\label{fig:scatter_abc_s1p5}
\end{figure}

A numerical search yields   the maximum violation for the correlation when 
$C_3=4.5,C_1=-3.5,C_2=C_0=0$, corresponding to the observable
\begin{equation}
O_A=\frac{9}{2}(\vec{S}\cdot \hat{a})^3-\frac{7}{2}(\vec{S}\cdot \hat{a)}
\end{equation}
For this optimal observable, the maximum violation is pegged at   $\mathcal{B}^{\frac{3}{2}}_3=2.45$,  with the violating region being $\sim$ 6.67\% of the total volume. We show the scatter of coefficients $C_1,C_2,C_3$ for constraint $C_0=0$  in Fig.~\ref{fig:scatter_abc_s1p5}.

It would be premature to draw any conclusion from the reduced relative region of contribution, as this is only a \emph{partial} contribution. Combined with the biquadratic contribution, we see that there is no \emph{significant} diminution in the over all area.

\subsubsection{$s \ge 2$}
\begin{figure}[H]
\includegraphics[width=0.5\textwidth]{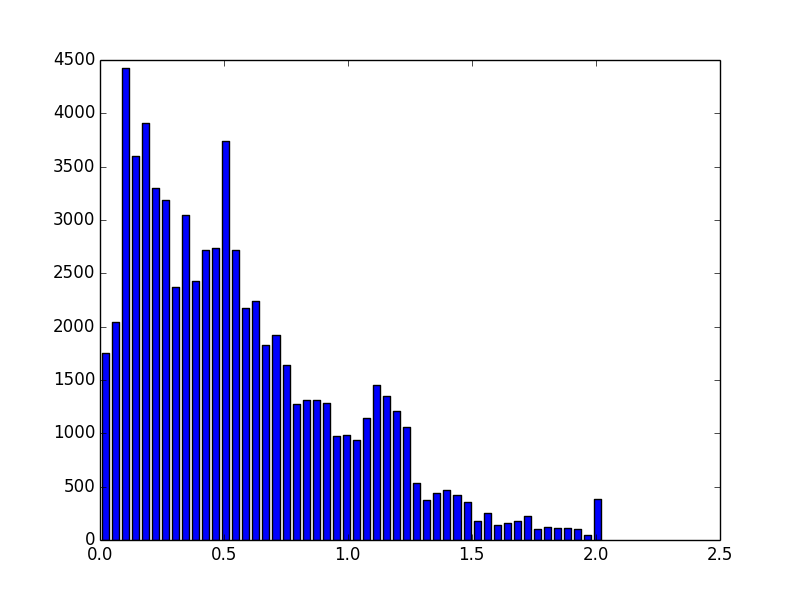}
\caption{Bell function $\mathcal{B}^2_3$ for all possible valid coefficient and configuration combinations.}
\label{fig:hist_bellcubic12_2}
\end{figure}

We conclude  the discussion on cubic correlations with a brief discussion of higher spins. Similarly to the quadratic case, $s=2$ shows a mild violation, with a maximum value  of the Bell function given by $\mathcal{B}^{2}_3=2.03$ for the configuration $C_3=4,C_1=-3,C_2=C_0=0$.  The histogram of this Bell function for all possible valid coefficient combinations with the standard planar experimental configuration is shown in  Fig.~\ref{fig:hist_bellcubic12_2}. The Bell inequality is respected for the cubic correlations involving all spins $s\geq\frac{5}{2}$.

\section{Biquartic correlations for $s =2$}\label{sect:4pole}
In this last example, we now consider a specific quartic observable for $s=2$:
\begin{equation}
O_A=2(\vec{\Sigma}\cdot \hat{a})^4-1
\end{equation}
which clearly spans the full range $[-1,+1]$. This observable is nontrivial for $s \ge 2$, and we restrict ourselves to spin 2 here. 
A straight forward evaluation yields the correlation to be
\begin{equation}
\mathcal{C}^2_4=\frac{7}{10}+\frac{45}{32}\sin^2\theta\cos^2\theta+\frac{39}{640}\sin^4\theta+\frac{51}{32}\cos^4\theta
\label{eq:quartic_corr}
\end{equation}

\begin{figure}[!htb]
\includegraphics[width=0.5\textwidth]{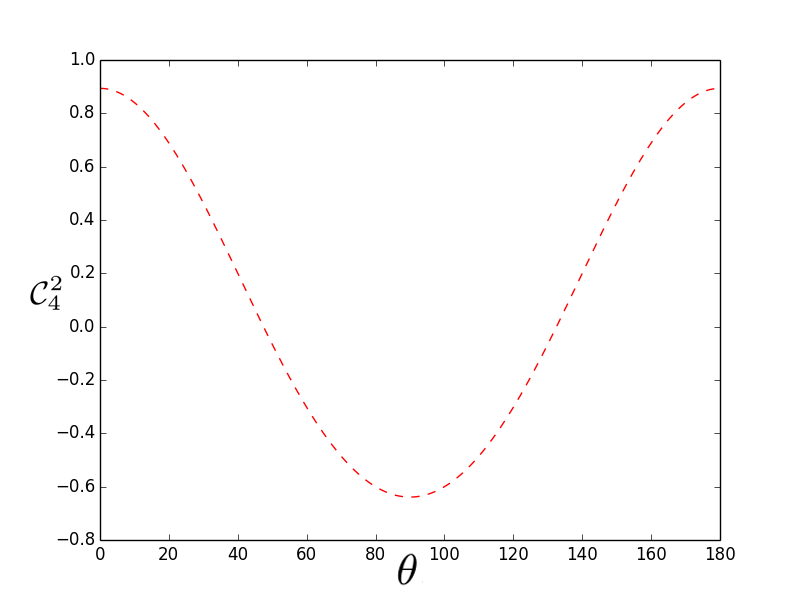}
\caption{The quartic correlator $\mathcal{C}^2_4$ for the $spin-2$ singlet state  given by Eq.~\ref{eq:quartic_corr}.}
\label{fig:fourpole_correlation_s2}
\end{figure}

\begin{figure}[!htb]
\includegraphics[width=0.5\textwidth]{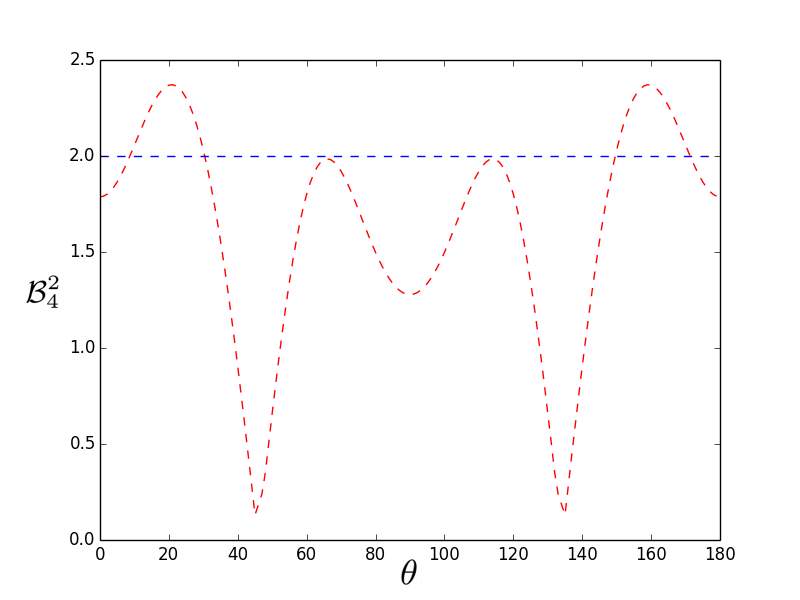}
\caption{Bell function $\mathcal{B}^2_4$, for the $spin-2$ singlet state from the correlator given in Eq.~\ref{eq:quartic_corr}. Blue: Bound of the inequality.}
\label{fig:bell_inequality_fourpole_corr_s2}
\end{figure}

Fig.~\ref{fig:fourpole_correlation_s2} plots this correlation.
The resulting Bell violation is shown in Fig.~\ref{fig:bell_inequality_fourpole_corr_s2}, and the maximum value attained by the Bell function is 
$\mathcal{B}^2_{4\max}=2.371$. The percentage area of violation is seen to be $\sim 2.5\%$. These signatures must be combined with the violation seen with the cubic correlation for a fuller picture.

\section{Summary  of results}\label{sect:results}
We first  summarize the results in a nutshell. The examples considered above suggest strongly that 
\begin{enumerate}
\item  Non-locality in $N$ level systems will be seen if the observables employed are of degree $N$,  with a small leeway for lower order observables. 
\item  The magnitude of violation does not see any significant diminution with increasing $N$. The relative region in the parameter space remains roughly constant if the observables are scaled in this manner. 
\item Finally, inspite of optimization, and in spite of employing the most nonlocal state, the violation fails to reach the maximum allowed value, $2\sqrt{2}$.
\end{enumerate}
 The results are collected  and displayed in table 1.
\begin{center}
\begin{table}[H]
\caption{Summary of Results}
\begin{tabular}{|c|c|c|c|c|} 
 \hline
 \textbf{$s$} & \textbf{$\mathcal{O}(O)$}  & \textbf{$O_A$}  & \textbf{$\mathcal{B}_\text{max}$} & $\textbf{\% Area of Violation}$  \\
 $\frac{1}{2}$  & 1 & $(\vec{S}\cdot \hat{a})$ & $2\sqrt{2}$  & $\approx 16.32\%$ \\
 1 & 2 & $2(\vec{S}\cdot \hat{a})^2-1$ & 2.55 & $\approx 14.7\%$ \\
 $\frac{3}{2}$ & 2 & $\frac{9}{4}(\vec{S}\cdot \hat{a})^2-\frac{5}{4}$ & 2.62 & $\approx 7.85\%$\\
  $\frac{3}{2}$ & 3 & $\frac{9}{2}(\vec{S}\cdot \hat{a})^3-\frac{7}{2}(\vec{S}\cdot \hat{a})$ & 2.45 & $\approx 6.67\%$\\
 2 & 3 & $4(\vec{S}\cdot \hat{a})^3-3(\vec{S}\cdot \hat{a})$ & 2.03 & $\approx 0.1\%$\\
 2 & 4 & $\frac{(\vec{S}\cdot \hat{a})^4}{8}-1$ & 2.371 & $\approx 2.522\%$\\
 \hline
\end{tabular}
\label{tab:summary}
\end{table}
\end{center}
\section{Discussion and comparison with other works}
Studies in non-locality have been dominated by three issues: (i) Detection of non-locality in $N$ dimensional systems, (ii) Extent of non-locality vis-a-vis the two qubit system, and by extension, (iii) Fate of non-locality in the large $N$ limit. The three issues are not unrelated to each other. 

One hurdle in comparing results  is in the different measures of non-locality employed, not only in different works, but also for different spins in the same work. Care must therefore be exercised since a na\"{i}ve  normalization may lead to misleading conclusions on the degree of non-locality. With this caveat, we first look at the last issue, the large $N$ limit, since it addresses the all important question of the classical limit of quantum systems.

\subsection{Large $N$ limit}
The fate of non-locality in the large $N$ limit was first studied by Mermin \cite{Merm80}, who concluded that 
non-locality persists in all dimensions. He argued that the classical limit is realized as an asymptote,   as $N \rightarrow \infty$. While the extent of violation was not seen to decrease, the violating region in the parameter space was seen to shrink to zero. This result seemed to gain support from a number of other works \cite{Arde91,Garg82,Merm80,Wodk94}. These conclusions were based on specific choices of correlations, and  it has been realized  to be erroneous  by Garg and Mermin \cite{Garg84}. Subsequently Ardehali \cite{Arde92}, and more recently  Dagomir et. al. \cite{Kasz00} have argued that there are correlations for which  non-locality, in fact, increase with the increasing dimension of the Hilbert space. Their results are, of course, not in conflict with the Cirel'son theorem \cite{Cirelson80} and the more general results obtained in \cite{Peres92,Gis92}.

Interesting that the above studies are, we observe that  --  bereft of any other criterion - neither the value nor the range of violation is, in fact, an issue for $N$ level systems. As observed by Bell in his seminal paper\cite{Bellinequalities},  any correlated state of two $N$ level systems can show as large a violation of Bell inequality as the two qubit system. Indeed, all that one needs to do is to project out a convenient four dimensional subspace spanned by a basis consisting of four separable pure states. The resultant quasi two qubit system would exhibit maximum violation, which cannot be improved upon further, thanks to Cirel'son's theorem \cite{Cirelson80}\footnote{That does not render experiments fruitless because any violation of Cirel'son's theorem would also negate quantum mechanics as a fundamental theory.}. Hence, non-locality in $N$ level systems is to be discussed with correlations that genuinely probe the full state space, and the large $N$ limit needs a rephrasing. The real interest, therefore, lies in viewing non-locality in $N$ level systems through the prism of  genuine qudit correlations which do not mimic any lower dimensional correlation. Equivalently, the eigenstates of observables in each $N$ system must span the full space.

With this condition, for the class of observables under consideration, we now demonstrate that the so called classical limit can indeed be achieved, but in a weaker sense. More precisely, we show that the so called classical limit is a manifestation of limited  experimental resources. 

\subsubsection{Weak classical limit}
Suppose that one has resources to measure upto $k^{th}$ order observables in spin space. Recall that for a given $s$, $k_{max} =2s$. We then consider the  constraints on a $k^{th}$ degree observable $O_A=\sum_{l=0}^{k} C_l (\vec{S}^s_A\cdot \hat{a})^l$ by the unit norm bound. We get $2N$ constraints:
\begin{equation}
-1 \le \sum_{l=0}^k C_l m^l  \le 1; -s \le m \le +s
\end{equation}
where $k \le 2s$. The dimension of the parameter space is $N_p =k+1$ which is to be compared with the number of bounds, $N_b= 2(2s+1)$. 

For a fixed order of the observable, as $s$ increases,  $N_b \gg N_p$, thanks to which one has to identify the region in the parameter space that is compatible with all the equations. The allowed region is the common intersection of regions culled out by each inequality. It is not difficult to see that as $s\rightarrow \infty$, the allowed region  shrinks rapidly to  the allowed to a point $C_l \rightarrow \pm \delta_{l,0}$. Bell inequality is trivially respected. 

On the other hand, suppose that the order of the polynomial function in spin scales with $s$. We would then have a polynomial of degree $2s$ subject to $2s+1$ equations. $N_b \sim N_p$ and for that reason   the allowed region in the parameter space does not shrink. The few simple examples that we discussed in the paper illustrate precisely that (see Table 1).

 \subsection{Extent of violation in $N$ level systems}
 We now turn our attention to compare our results with some of the earlier works. In particular, we now focus on the magnitude of violation. Unfortunately, almost none of the earlier works give information on the range of parameters over which violation occurs. Several measures employ inequalities which depend on $N$, which further makes it difficult to make meaningful comparisons.  
Table 2 presents  the results obtained  in \cite{Peres92, Bana98,Gis92} since all of them use the Bell inequality formulation employed in this paper. Note: the work reported in \cite{Bana98} admits a comparison only when $N \rightarrow \infty$.

Table 2 gives a summary of the results obtained in \cite{Peres92, Bana98,Gis92}, for $s=1$. Note that the observables in each of the works are quite different from each other and a comparison is, therefore, of quite some interest. So are the measures of non-locality employed. The results in table 2 must be, therefore, interpreted carefully.

With the approach given by Peres \cite{Peres92}, while the magnitude of the violation tends to a constant (a claim consistent with that of Banaszak \cite{Bana98}), the range of parameters for which a violation is detected becomes vanishingly small as $s\rightarrow \infty$.  Contrarily, our results indicate that once a suitable dimension of parameter space is established, said region of violation consists of a set of orthogonal subspaces that do not diminish asymptotically.  Finding the right observables of a suitable order dependent on $s$ is the key. 

At this juncture, it is pertinent to note that Peres \cite{Peres92} also makes a note of experimental limitations on observing non-locality: provided that consecutive $m$'s are distinguishable, non-locality \emph{should} be observable. This goes hand-in-hand with our notion of a `weak' classical limit being attained due to limited experimental resources. This argument is strengthened by that of Gisin \cite{Gis92}, who shows that mere selection of large quantum numbers does not necessarily indicate classical behaviour.  Also provided that one can construct pairs of observables for which violation occurs, large quantum numbers are of no consequence.  The present study, apart from being  in consonance with these observations, brings out the need for  those pairs of observables to be nontrivial. They should not be defined in a proper subspace and they must be of sufficiently high rank.

\begin{center}
\begin{table}
\caption{Summary of Results}
 \resizebox{\columnwidth}{!}{%
\begin{tabular}{|c|c|c|c|}

 \hline
 \textbf{Author} & \textbf{Inequality}  & \textbf{Behaviour $s (r)\rightarrow\infty$} & \textbf{Violation} \\
Peres & $f(C)\leq 2$ & $f(C)\rightarrow \frac{3\sin x}{x}-\frac{\sin3x}{3x}$ & 2.55 for $s=1$\\
Gisin  & $f(j) \leq 2$ & $f(j)\rightarrow 2\sqrt{2}(1-\frac{0.1464}{s}$) & 2.55 for $s=1$ \\
Banaszek & $|\mathcal{B}|\leq 2$ & $\mathcal{B}\approx 2.19$ & 2.19 $\forall$ EPR states\\
 \hline
\end{tabular}
\label{tab:compare}
}
\end{table}
\end{center}

\section{Non-locality and geometry of spin states}
A common feature among  all  studies reporting  violation of Bell inequality is  that none of them attain the maximum violation that is allowed by the Cirel'son bound \cite{Cirelson80}. It becomes all the more striking if we remember that there are observables defined in four dimensional subspaces whose correlations are as good as --- in fact equivalent to -- the two qubit case.  The reason for this failure merits an answer, which lies in a more careful analysis of the geometry of spin states.

States of a spin half particle admit a representation on the Bloch sphere, which essentially means that they can always be looked upon as possessing a definite value of $S_z$ along a suitable quantization axis: $\vert \psi \rangle \equiv \vert m=\frac{1}{2}\rangle_{\hat{a}}$ where $\hat{a}$ is the quantization axis. This does not hold for higher spins and the set of states that can be identified as belonging to an eigenvalue of $S_z$ with respect to any quantization axis is a set of measure zero. In other words, the state of a particle with  a spin $ s \ge 1$ does not, in general admit a Bloch sphere representation. Consequently, the observables and the correlations that we have considered in this paper are not generic in nature.

The most general representation of a pure spin $s$ state is multivector in nature. Indeed, as shown in
\cite{Majo32, Ravi87}, the state (pure) of a spin $s$ particle can be represented by $2s$ points on a sphere. If the particle is in a state $\vert s,m\rangle$, then the description collapses to $s+m$ identical points and diametrically opposite $s-m$ identical points, along the quantization axis, on the sphere. This means that in general, an observable for a spin $s$ particle involves $2s$ independent directions, and consequently, a correlation for a coupled system of two spin $S$ particles involves $4s$ independent directions. Coupled with the fact that a pure state is determined almost completely -- except for discrete ambiguities -- by its highest rank observable, {\it viz}, it is not surprising that we need observables of order $2s$. Correlations involving such observables exhaust all the possibilities and should, therefore, yield maximal violation. However, this makes the theoretical analysis - and its experimental implementation -- more complicated. We address this problem elsewhere.
\section{Conclusion}\label{sect:conclusion}
In conclusion, we have undertaken a comprehensive study of non-locality in coupled high spin systems with emphasis on (i) suitable observables that display  non-locality,  (ii) the so called classical limit when $N=(2s+1) \rightarrow \infty$, (iii) genuine observables that span the Hilbert space and also exhibit non-locality, and finally (iv) the need to go beyond the standard Stern-Gerlach set ups and observables that are tied to bivectors in the two spin systems. Hopefully, this study will spur further activity experimentally to detect and characterize  non-locality in spin systems.

\bibliography{bibliography}

\begin{thebibliography}{24}%
\makeatletter
\providecommand \@ifxundefined [1]{%
 \@ifx{#1\undefined}
}%
\providecommand \@ifnum [1]{%
 \ifnum #1\expandafter \@firstoftwo
 \else \expandafter \@secondoftwo
 \fi
}%
\providecommand \@ifx [1]{%
 \ifx #1\expandafter \@firstoftwo
 \else \expandafter \@secondoftwo
 \fi
}%
\providecommand \natexlab [1]{#1}%
\providecommand \enquote  [1]{``#1''}%
\providecommand \bibnamefont  [1]{#1}%
\providecommand \bibfnamefont [1]{#1}%
\providecommand \citenamefont [1]{#1}%
\providecommand \href@noop [0]{\@secondoftwo}%
\providecommand \href [0]{\begingroup \@sanitize@url \@href}%
\providecommand \@href[1]{\@@startlink{#1}\@@href}%
\providecommand \@@href[1]{\endgroup#1\@@endlink}%
\providecommand \@sanitize@url [0]{\catcode `\\12\catcode `\$12\catcode
  `\&12\catcode `\#12\catcode `\^12\catcode `\_12\catcode `\%12\relax}%
\providecommand \@@startlink[1]{}%
\providecommand \@@endlink[0]{}%
\providecommand \url  [0]{\begingroup\@sanitize@url \@url }%
\providecommand \@url [1]{\endgroup\@href {#1}{\urlprefix }}%
\providecommand \urlprefix  [0]{URL }%
\providecommand \Eprint [0]{\href }%
\providecommand \doibase [0]{http://dx.doi.org/}%
\providecommand \selectlanguage [0]{\@gobble}%
\providecommand \bibinfo  [0]{\@secondoftwo}%
\providecommand \bibfield  [0]{\@secondoftwo}%
\providecommand \translation [1]{[#1]}%
\providecommand \BibitemOpen [0]{}%
\providecommand \bibitemStop [0]{}%
\providecommand \bibitemNoStop [0]{.\EOS\space}%
\providecommand \EOS [0]{\spacefactor3000\relax}%
\providecommand \BibitemShut  [1]{\csname bibitem#1\endcsname}%
\let\auto@bib@innerbib\@empty
\bibitem [{\citenamefont {Bell}(1964)}]{Bellinequalities}%
  \BibitemOpen
  \bibfield  {author} {\bibinfo {author} {\bibfnamefont {J.~S.}\ \bibnamefont
  {Bell}},\ }\href@noop {} {\bibfield  {journal} {\bibinfo  {journal}
  {Physics}\ }\textbf {\bibinfo {volume} {1}},\ \bibinfo {pages} {195}
  (\bibinfo {year} {1964})}\BibitemShut {NoStop}%
\bibitem [{\citenamefont {Mermin}(1980)}]{Merm80}%
  \BibitemOpen
  \bibfield  {author} {\bibinfo {author} {\bibfnamefont {N.~D.}\ \bibnamefont
  {Mermin}},\ }\href@noop {} {\bibfield  {journal} {\bibinfo  {journal} {Phys.
  Rev. D}\ }\textbf {\bibinfo {volume} {22}},\ \bibinfo {pages} {356} (\bibinfo
  {year} {1980})}\BibitemShut {NoStop}%
\bibitem [{\citenamefont {Kaszlikowski}\ \emph {et~al.}(2000)\citenamefont
  {Kaszlikowski}, \citenamefont {Gnaci\ifmmode~\acute{n}\else \'{n}\fi{}ski},
  \citenamefont {\ifmmode~\dot{Z}\else \.{Z}\fi{}ukowski}, \citenamefont
  {Miklaszewski},\ and\ \citenamefont {Zeilinger}}]{Kasz00}%
  \BibitemOpen
  \bibfield  {author} {\bibinfo {author} {\bibfnamefont {D.}~\bibnamefont
  {Kaszlikowski}}, \bibinfo {author} {\bibfnamefont {P.}~\bibnamefont
  {Gnaci\ifmmode~\acute{n}\else \'{n}\fi{}ski}}, \bibinfo {author}
  {\bibfnamefont {M.}~\bibnamefont {\ifmmode~\dot{Z}\else \.{Z}\fi{}ukowski}},
  \bibinfo {author} {\bibfnamefont {W.}~\bibnamefont {Miklaszewski}}, \ and\
  \bibinfo {author} {\bibfnamefont {A.}~\bibnamefont {Zeilinger}},\ }\href
  {\doibase 10.1103/PhysRevLett.85.4418} {\bibfield  {journal} {\bibinfo
  {journal} {Phys. Rev. Lett.}\ }\textbf {\bibinfo {volume} {85}},\ \bibinfo
  {pages} {4418} (\bibinfo {year} {2000})}\BibitemShut {NoStop}%
\bibitem [{\citenamefont {Garg}\ and\ \citenamefont {Mermin}(1984)}]{Garg84}%
  \BibitemOpen
  \bibfield  {author} {\bibinfo {author} {\bibfnamefont {A.}~\bibnamefont
  {Garg}}\ and\ \bibinfo {author} {\bibfnamefont {N.}~\bibnamefont {Mermin}},\
  }\href {\doibase 10.1007/BF00741645} {\bibfield  {journal} {\bibinfo
  {journal} {Foundations of Physics}\ }\textbf {\bibinfo {volume} {14}},\
  \bibinfo {pages} {1} (\bibinfo {year} {1984})}\BibitemShut {NoStop}%
\bibitem [{\citenamefont {Peres}(1992)}]{Peres92}%
  \BibitemOpen
  \bibfield  {author} {\bibinfo {author} {\bibfnamefont {A.}~\bibnamefont
  {Peres}},\ }\href {\doibase 10.1103/PhysRevA.46.4413} {\bibfield  {journal}
  {\bibinfo  {journal} {Phys. Rev. A}\ }\textbf {\bibinfo {volume} {46}},\
  \bibinfo {pages} {4413} (\bibinfo {year} {1992})}\BibitemShut {NoStop}%
\bibitem [{\citenamefont {Braunstein}\ \emph {et~al.}(1992)\citenamefont
  {Braunstein}, \citenamefont {Mann},\ and\ \citenamefont {Revzen}}]{Brau92}%
  \BibitemOpen
  \bibfield  {author} {\bibinfo {author} {\bibfnamefont {S.}~\bibnamefont
  {Braunstein}}, \bibinfo {author} {\bibfnamefont {A.}~\bibnamefont {Mann}}, \
  and\ \bibinfo {author} {\bibfnamefont {M.}~\bibnamefont {Revzen}},\ }\href
  {\doibase 10.1103/PhysRevLett.68.3259} {\bibfield  {journal} {\bibinfo
  {journal} {Phys. Rev. Lett.}\ }\textbf {\bibinfo {volume} {68}},\ \bibinfo
  {pages} {3259} (\bibinfo {year} {1992})}\BibitemShut {NoStop}%
\bibitem [{\citenamefont {Gisin}\ and\ \citenamefont {Peres}(1992)}]{Gis92}%
  \BibitemOpen
  \bibfield  {author} {\bibinfo {author} {\bibfnamefont {N.}~\bibnamefont
  {Gisin}}\ and\ \bibinfo {author} {\bibfnamefont {A.}~\bibnamefont {Peres}},\
  }\href@noop {} {\bibfield  {journal} {\bibinfo  {journal} {Phys. Lett. A}\
  }\textbf {\bibinfo {volume} {162}},\ \bibinfo {pages} {15} (\bibinfo {year}
  {1992})}\BibitemShut {NoStop}%
\bibitem [{\citenamefont {Collins}\ \emph
  {et~al.}(2002{\natexlab{a}})\citenamefont {Collins}, \citenamefont {Gisin},
  \citenamefont {Linden}, \citenamefont {Massar},\ and\ \citenamefont
  {Popescu}}]{Coll02}%
  \BibitemOpen
  \bibfield  {author} {\bibinfo {author} {\bibfnamefont {D.}~\bibnamefont
  {Collins}}, \bibinfo {author} {\bibfnamefont {N.}~\bibnamefont {Gisin}},
  \bibinfo {author} {\bibfnamefont {N.}~\bibnamefont {Linden}}, \bibinfo
  {author} {\bibfnamefont {S.}~\bibnamefont {Massar}}, \ and\ \bibinfo {author}
  {\bibfnamefont {S.}~\bibnamefont {Popescu}},\ }\href {\doibase
  10.1103/PhysRevLett.88.040404} {\bibfield  {journal} {\bibinfo  {journal}
  {Phys. Rev. Lett.}\ }\textbf {\bibinfo {volume} {88}},\ \bibinfo {pages}
  {040404} (\bibinfo {year} {2002}{\natexlab{a}})}\BibitemShut {NoStop}%
\bibitem [{\citenamefont {W\'odkiewicz}(1995)}]{Wodk95}%
  \BibitemOpen
  \bibfield  {author} {\bibinfo {author} {\bibfnamefont {K.}~\bibnamefont
  {W\'odkiewicz}},\ }\href@noop {} {\bibfield  {journal} {\bibinfo  {journal}
  {Phys. Rev. A}\ }\textbf {\bibinfo {volume} {52}} (\bibinfo {year}
  {1995})}\BibitemShut {NoStop}%
\bibitem [{\citenamefont {Wu}\ \emph {et~al.}(2001)\citenamefont {Wu},
  \citenamefont {Zong}, \citenamefont {Pang},\ and\ \citenamefont
  {Wang}}]{Wu01}%
  \BibitemOpen
  \bibfield  {author} {\bibinfo {author} {\bibfnamefont {X.}~\bibnamefont
  {Wu}}, \bibinfo {author} {\bibfnamefont {H.}~\bibnamefont {Zong}}, \bibinfo
  {author} {\bibfnamefont {H.}~\bibnamefont {Pang}}, \ and\ \bibinfo {author}
  {\bibfnamefont {F.}~\bibnamefont {Wang}},\ }\href {\doibase
  http://dx.doi.org/10.1016/S0375-9601(01)00095-0} {\bibfield  {journal}
  {\bibinfo  {journal} {Physics Letters A}\ }\textbf {\bibinfo {volume}
  {281}},\ \bibinfo {pages} {203 } (\bibinfo {year} {2001})}\BibitemShut
  {NoStop}%
\bibitem [{\citenamefont {Fu}(2004)}]{FuLi04}%
  \BibitemOpen
  \bibfield  {author} {\bibinfo {author} {\bibfnamefont {L.}~\bibnamefont
  {Fu}},\ }\href {\doibase 10.1103/PhysRevLett.92.130404} {\bibfield  {journal}
  {\bibinfo  {journal} {Phys. Rev. Lett.}\ }\textbf {\bibinfo {volume} {92}},\
  \bibinfo {pages} {130404} (\bibinfo {year} {2004})}\BibitemShut {NoStop}%
\bibitem [{\citenamefont {Cirel'son}(1980)}]{Cirelson80}%
  \BibitemOpen
  \bibfield  {author} {\bibinfo {author} {\bibfnamefont {B.}~\bibnamefont
  {Cirel'son}},\ }\href {\doibase 10.1007/BF00417500} {\bibfield  {journal}
  {\bibinfo  {journal} {Lett.Math.Phys.}\ }\textbf {\bibinfo {volume} {4}},\
  \bibinfo {pages} {93} (\bibinfo {year} {1980})}\BibitemShut {NoStop}%
\bibitem [{\citenamefont {Collins}\ \emph
  {et~al.}(2002{\natexlab{b}})\citenamefont {Collins}, \citenamefont {Gisin},
  \citenamefont {Popescu}, \citenamefont {Roberts},\ and\ \citenamefont
  {Scarani}}]{Coll02_2}%
  \BibitemOpen
  \bibfield  {author} {\bibinfo {author} {\bibfnamefont {D.}~\bibnamefont
  {Collins}}, \bibinfo {author} {\bibfnamefont {N.}~\bibnamefont {Gisin}},
  \bibinfo {author} {\bibfnamefont {S.}~\bibnamefont {Popescu}}, \bibinfo
  {author} {\bibfnamefont {D.}~\bibnamefont {Roberts}}, \ and\ \bibinfo
  {author} {\bibfnamefont {V.}~\bibnamefont {Scarani}},\ }\href {\doibase
  10.1103/PhysRevLett.88.170405} {\bibfield  {journal} {\bibinfo  {journal}
  {Phys. Rev. Lett.}\ }\textbf {\bibinfo {volume} {88}},\ \bibinfo {pages}
  {170405} (\bibinfo {year} {2002}{\natexlab{b}})}\BibitemShut {NoStop}%
\bibitem [{\citenamefont {Mitchell}\ \emph {et~al.}(2004)\citenamefont
  {Mitchell}, \citenamefont {Popescu},\ and\ \citenamefont
  {Roberts}}]{Mitch04}%
  \BibitemOpen
  \bibfield  {author} {\bibinfo {author} {\bibfnamefont {P.}~\bibnamefont
  {Mitchell}}, \bibinfo {author} {\bibfnamefont {S.}~\bibnamefont {Popescu}}, \
  and\ \bibinfo {author} {\bibfnamefont {D.}~\bibnamefont {Roberts}},\ }\href
  {\doibase 10.1103/PhysRevA.70.060101} {\bibfield  {journal} {\bibinfo
  {journal} {Phys. Rev. A}\ }\textbf {\bibinfo {volume} {70}},\ \bibinfo
  {pages} {060101} (\bibinfo {year} {2004})}\BibitemShut {NoStop}%
\bibitem [{Note1()}]{Note1}%
  \BibitemOpen
  \bibinfo {note} {More generally, one could consider correlations of degree
  $k~(k')$ in subsystem $A(B)$, but they are of no consequence to us
  here.}\BibitemShut {Stop}%
\bibitem [{Note2()}]{Note2}%
  \BibitemOpen
  \bibinfo {note} {The numerical results agree with the analytic results which
  are easily available for $s=\protect \frac {1}{2},~\protect \frac
  {3}{2}$.}\BibitemShut {Stop}%
\bibitem [{\citenamefont {Banaszek}\ and\ \citenamefont
  {W\'odkiewicz}(1998)}]{Bana98}%
  \BibitemOpen
  \bibfield  {author} {\bibinfo {author} {\bibfnamefont {K.}~\bibnamefont
  {Banaszek}}\ and\ \bibinfo {author} {\bibfnamefont {K.}~\bibnamefont
  {W\'odkiewicz}},\ }\href {\doibase 10.1103/PhysRevA.58.4345} {\bibfield
  {journal} {\bibinfo  {journal} {Phys. Rev. A}\ }\textbf {\bibinfo {volume}
  {58}},\ \bibinfo {pages} {4345} (\bibinfo {year} {1998})}\BibitemShut
  {NoStop}%
\bibitem [{\citenamefont {Ardehali}(1991)}]{Arde91}%
  \BibitemOpen
  \bibfield  {author} {\bibinfo {author} {\bibfnamefont {M.}~\bibnamefont
  {Ardehali}},\ }\href {\doibase 10.1103/PhysRevD.44.3336} {\bibfield
  {journal} {\bibinfo  {journal} {Phys. Rev. D}\ }\textbf {\bibinfo {volume}
  {44}},\ \bibinfo {pages} {3336} (\bibinfo {year} {1991})}\BibitemShut
  {NoStop}%
\bibitem [{\citenamefont {Garg}\ and\ \citenamefont {Mermin}(1982)}]{Garg82}%
  \BibitemOpen
  \bibfield  {author} {\bibinfo {author} {\bibfnamefont {A.}~\bibnamefont
  {Garg}}\ and\ \bibinfo {author} {\bibfnamefont {N.~D.}\ \bibnamefont
  {Mermin}},\ }\href {\doibase 10.1103/PhysRevLett.49.901} {\bibfield
  {journal} {\bibinfo  {journal} {Phys. Rev. Lett.}\ }\textbf {\bibinfo
  {volume} {49}},\ \bibinfo {pages} {901} (\bibinfo {year} {1982})}\BibitemShut
  {NoStop}%
\bibitem [{\citenamefont {Scully}\ and\ \citenamefont
  {W\'odkiewicz}(1994)}]{Wodk94}%
  \BibitemOpen
  \bibfield  {author} {\bibinfo {author} {\bibfnamefont {M.}~\bibnamefont
  {Scully}}\ and\ \bibinfo {author} {\bibfnamefont {K.}~\bibnamefont
  {W\'odkiewicz}},\ }\href {\doibase 10.1007/BF02053909} {\bibfield  {journal}
  {\bibinfo  {journal} {Foundations of Physics}\ }\textbf {\bibinfo {volume}
  {24}},\ \bibinfo {pages} {85} (\bibinfo {year} {1994})}\BibitemShut {NoStop}%
\bibitem [{\citenamefont {Ardehali}(1992)}]{Arde92}%
  \BibitemOpen
  \bibfield  {author} {\bibinfo {author} {\bibfnamefont {M.}~\bibnamefont
  {Ardehali}},\ }\href {\doibase 10.1103/PhysRevA.46.5375} {\bibfield
  {journal} {\bibinfo  {journal} {Phys. Rev. A}\ }\textbf {\bibinfo {volume}
  {46}},\ \bibinfo {pages} {5375} (\bibinfo {year} {1992})}\BibitemShut
  {NoStop}%
\bibitem [{Note3()}]{Note3}%
  \BibitemOpen
  \bibinfo {note} {That does not render experiments fruitless because any
  violation of Cirelson's theorem would also negate quantum mechanics as a
  fundamental theory.}\BibitemShut {Stop}%
\bibitem [{\citenamefont {Majorana}(1932)}]{Majo32}%
  \BibitemOpen
  \bibfield  {author} {\bibinfo {author} {\bibfnamefont {E.}~\bibnamefont
  {Majorana}},\ }\href@noop {} {\bibfield  {journal} {\bibinfo  {journal} {Il
  Nuovo Cimento (1924-1942)}\ }\textbf {\bibinfo {volume} {9}},\ \bibinfo
  {pages} {43} (\bibinfo {year} {1932})}\BibitemShut {NoStop}%
\bibitem [{\citenamefont {Ravishankar}\ and\ \citenamefont
  {Ramachandran}(1987)}]{Ravi87}%
  \BibitemOpen
  \bibfield  {author} {\bibinfo {author} {\bibfnamefont {V.}~\bibnamefont
  {Ravishankar}}\ and\ \bibinfo {author} {\bibfnamefont {G.}~\bibnamefont
  {Ramachandran}},\ }\href {\doibase 10.1103/PhysRevC.35.62} {\bibfield
  {journal} {\bibinfo  {journal} {Phys. Rev. C}\ }\textbf {\bibinfo {volume}
  {35}},\ \bibinfo {pages} {62} (\bibinfo {year} {1987})}\BibitemShut {NoStop}%
\end{thebibliography}%

\end{document}